\documentclass[pra,aps,twocolumn,superscriptaddress,showpacs,10pt]{revtex4-1}

\usepackage[FIGTOPCAP,tight]{subfigure} 
\usepackage{amsmath}
\usepackage{amssymb}
\usepackage{amsthm}
\usepackage{float}
\usepackage[T1]{fontenc}
\usepackage[latin9]{inputenc}
\usepackage{geometry}
\usepackage{graphicx}
\usepackage{esint}
\usepackage{array}
\usepackage{multirow}
\usepackage{xcolor}
\usepackage[normalem]{ulem}
\usepackage{lipsum}
\usepackage{dcolumn}

\usepackage{mathrsfs}

\usepackage{mathtools}

\usepackage{float}

\usepackage{bm}

\usepackage{mathrsfs}

\usepackage{hyperref}

\newcommand{\ket}[1]{\left|#1\right\rangle}
\newcommand{\bra}[1]{\left\langle#1\right|}

\newcommand{\mbf}[1]{\mathbf{#1}}

\usepackage{verbatim} 

\usepackage{comment}

\newcommand{\be}{\begin{equation}}
\newcommand{\bea}{\begin{eqnarray}}
\newcommand{\ee}{\end{equation}}
\newcommand{\eea}{\end{eqnarray}}
\newcommand{\ben}{\begin{equation*}}
\newcommand{\bean}{\begin{eqnarray*}}
\newcommand{\een}{\end{equation*}}
\newcommand{\eean}{\end{eqnarray*}}
\newcommand{\ba}{\begin{align}}
\newcommand{\ea}{\end{align}}
\newcommand{\ban}{\begin{align*}}
\newcommand{\ean}{\end{align*}}

\newcommand{\tcr}[1]{\textcolor{red}{#1}}

\graphicspath{{./Images/}}

\begin{document}

\title{Quantum imaging using relativistic detectors}

\author{Nicholas Bornman}
\affiliation{School of Physics, University of the Witwatersrand, Johannesburg 2000, South Africa}

\author{Achim Kempf}
\affiliation{Department of Applied Mathematics, University of Waterloo, Waterloo, Ontario, Canada N2L 3G1}
\affiliation{Institute for Quantum Computing, University of Waterloo, Waterloo, Ontario, Canada N2L 3G1}
\affiliation{Perimeter Institute for Theoretical Physics, Waterloo, Ontario, Canada N2L 2Y5}

\author{Andrew Forbes}
\affiliation{School of Physics, University of the Witwatersrand, Johannesburg 2000, South Africa}

\date{\today}

\begin{abstract}

Imaging in quantum optics (QO) is usually formulated in the languages of quantum mechanics and Fourier optics. While relatively advanced fields, notions such as different reference frames and the degradation of entanglement due to acceleration do not usually feature. Here we propose the idea of using so-called Unruh-DeWitt (UDW) detectors to model the imaging process in QO. In particular, we first present a quantum field theory version of a state describing Spontaneous Parametric Down Conversion (SPDC), one of the principal processes employed to create entangled photons in the laboratory. This state, coupled to UDW detectors, is used to investigate a single-pixel ghost image under both inertial and non-inertial settings, and a two-pixel image under inertial conditions. The reconstructed images obtained for various possible inputs can be distinguished better than a pure guess, hence the formalism can be used to describe imaging between non-inertial frames. We briefly consider the origin of the correlations between the UDW detectors, which don't appear to arise from the usual notion of entanglement. Finally, we find that the contrast between the possible outcomes in the single-pixel case follows a curious coupling time dependent behaviour.

\end{abstract}

\pacs{}

\maketitle

\section{Introduction}
\label{sec:intro}

How to quantitatively describe the process of measurement in quantum physics has vexed scientists for many years: experimental readouts take place on the classical, macroscopic scale, yet the active elements of detectors generally operate on the quantum level. ``How, then, can one establish a correspondence between the quantum and the familiar classical reality?'' \cite{RevModPhys.75.715}. In addition to understanding the detection process in inertial frames, the inevitable rise of quantum-based communication between satellites, for example, heightens the need to understand the effects of non-inertial frames on the behaviour of entanglement and the detection process in practice. Indeed, quantum entanglement does not appear to be invariant under acceleration \cite{PhysRevA.79.052109}, and given that the very notion of a state's particle content (and therefore information content) is frame dependent \cite{PhysRevD.14.870}, relativity cannot strictly be ignored.

A popular model of particle detectors, which provide such a correspondence, was introduced by Unruh and DeWitt \cite{PhysRevD.14.870, hawking2010general} and consists of a first-quantized, two-level system (such as a two-level atom or the spin degree of freedom of an electron) coupled to the quantum field to be probed. The Unruh-DeWitt (UDW) detector has since been widely applied to, for example, the interaction between atoms and the electromagnetic field in cavities \cite{Lopp_2018}, the Casimir-Polder force \cite{PhysRevA.89.033835}, and also attempts to solve the issue of the non-existence of a position operator in QFT \cite{PhysRevA.89.042111}.

\begin{figure}[H]
\centering
\includegraphics[width=0.79\linewidth]{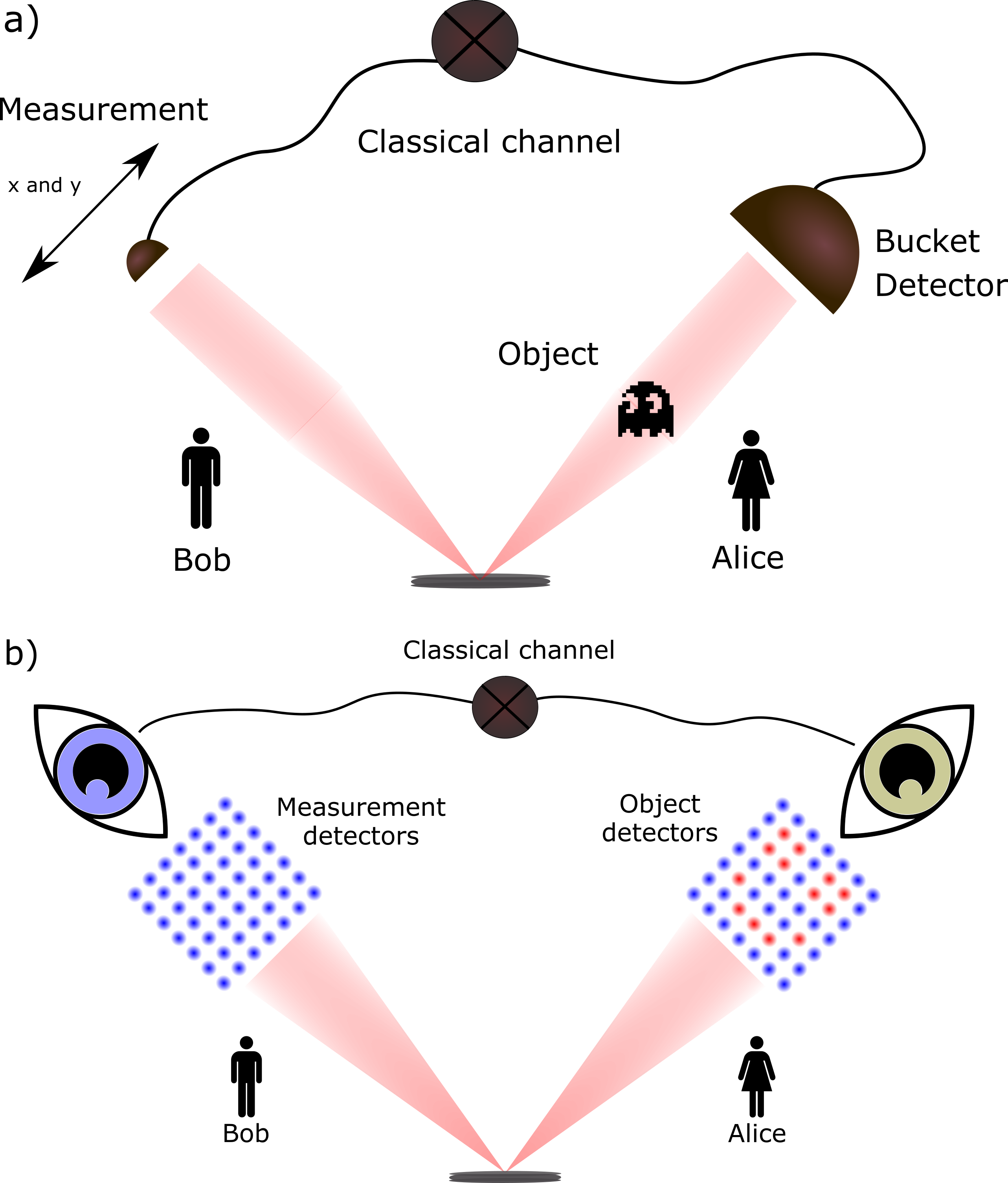}
\caption{a) Schematic of typical ghost imaging experiment; b) UDW imaging analogy}
\label{fig:schematic}
\end{figure}

In a quantum optics laboratory, important experiments such as quantum ghost imaging \cite{Shapiro2012} and observations of Hong-Ou-Mandel filtration \cite{PhysRevLett.59.2044} necessitate the detection of time-stamped and/or position-stamped entangled photons. A typical quantum ghost imaging \cite{PhysRevA.52.R3429, Shapiro2012} experiment entails pumping a non-linear crystal with coherent photons from a laser resulting in the creation of a pair of entangled photons by way of a process known as Spontaneous Parametric Down Conversion (SPDC) \cite{PhysRevLett.25.84}. After separating these photons into two paths, one photon from the pair interacts with a static object in one path and is thereafter registered by a detector which lacks spatial resolution; measurements are performed on the second photon using a spatially-resolved detector, Fig. \ref{fig:schematic} a). These measurements, along with classical information after post-selecting on the state of the first photon, allows the image to be reconstructed.

However, such studies do not take into account relativity. UDW detectors, which can account for relativity, fit well within the context of quantum optics. Indeed, it is argued in \cite{PhysRevD.92.104019} that models of the interaction between light and atoms employed in quantum optics, such as the Jaynes-Cummings model \cite{1443594, JPhysBAndrew}, are in fact derivatives of the UDW detector model.

This letter presents a novel model of the imaging process using UDW detectors, with particular focus on a relatively easily-implementable quantum optics experiment: ghost imaging. Inspired by the binary nature of UDW detectors, the purpose of this note is three-fold. Firstly, we briefly present a quantum field theory version of an SPDC state. Secondly, using this state, we then explore the potential of UDW detectors in explaining traditional quantum ghost imaging (in an inertial frame) using an analogy: each pixel the static object is comprised of, and the pixels of the device that is used to perform measurements on the second photon (such as a Spatial Light Modulator \cite{rosales2017shape}), are taken to be separate UDW detectors, Fig. \ref{fig:schematic} b). These detectors are prepared in some initial state and coupled to a field in the SPDC state. Finally, we consider the case of uniform acceleration of some of the detectors, and the effect this acceleration has on the reconstructed image.

\section{Unruh-DeWitt detector basics}
\label{sec:udwbasics}

We adopt the conventions of \cite{peskin1995introduction}, work in $d+1$ dimensional spacetime, and model the electromagnetic field as a scalar field $\hat{\phi}(t,\mathbf{x}) = \hat{\phi}(x)$ in the Dirac picture

\be
\hat{\phi}(x) = \int\frac{d^d k}{(2\pi)^d}\frac{1}{\sqrt{2 E_{\mbf{k}}}}\left( \hat{a}_{\mbf{k}}e^{-ik\cdot x} + \hat{a}^{\dagger}_{\mbf{k}}e^{ik \cdot x} \right),
\label{eqn:phi}
\ee

\noindent
where the usual relativistic dispersion relation holds

\be
E_{\mbf{k}} = \sqrt{|\mbf{k}|^2 + m^2}.
\ee

We model a quantum ghost imaging experiment by coupling a system of $n$ Unruh-DeWitt detectors to $\hat{\phi}$. Given an appropriate parameter (the proper time of the laboratory frame, for instance), the worldline $X_i^{\mu}$ of the centre of mass of detector $i$ can be parametrized. Each detector is given a finite spatial `smearing' profile, $f_i(\mbf{x})$, to avoid complications arising from regularizing point-like detectors; detector $i$ is coupled with $\hat{\phi}$ by way of its monopole moment operator $\hat{Q}_i(t)$ (assuming minimal coupling \cite{PhysRevA.89.033835})

\be
\hat{Q}_i(t) = e^{i\Omega_it}\ket{e}_i\bra{g}_i + e^{-i\Omega_it}\ket{g}_i\bra{e}_i,
\label{eqn:monopole}
\ee

\noindent
where $\ket{g}_i$ and $\ket{e}_i$ represent the ground/excited state of detector $i$, with $\Omega_i$ being the energy gap between the two levels. Detector $i$'s temporal behaviour (i.e. how it is switched `on' and `off') is given by the switching function $\varepsilon_i(t)$, and is coupled to $\hat{\phi}$ with coupling strength $\lambda_i$. The system's dynamics is then fully described by the interaction part of the Hamiltonian

\be
\hat{H}^{int}(t) = \sum_{i=1}^n \hat{H}^{int}_i(t),
\label{eqn:Hint}
\ee

\noindent
where

\be
\hat{H}^{int}_i(t) = \lambda_i \varepsilon_i(t) \hat{Q}_i(t) \int d^d x f_i(\mbf{x} - \mbf{X}_i(t))\hat{\phi}(x).
\label{eqn:Hinteach}
\ee

Next, denote the initial state of the $n$ detectors by $\ket{\mbf{E}}$ and $\ket{\mbf{K}}$ as the initial state of $\hat{\phi}$. Assuming the detectors and the field to be initially uncoupled, $\ket{\Psi(t_0)} = \ket{\mbf{E}} \otimes \ket{\mbf{K}}$ at some reference time $t_0$, the evolution of the system to $\ket{\Psi(t)} = \hat{U}(t,t_0)\ket{\Psi(t_0)}$ at time $t > t_0$ is determined by the \textit{Dyson operator}

\be
\hat{U}(t,t_0) = \mathcal{T} \; \text{exp}\left( -i \int_{t_0}^t \hat{H}^{int}(t')dt' \right),
\label{eqn:U}
\ee

\noindent
with $\mathcal{T}$ the \textit{time-ordering} operator. Without loss of generality, let $t_0 \to -\infty$ and $t \to \infty$, since all time dependence can be accommodated by suitable switching functions $\varepsilon_j$. Hence

\be
\hat{U} = \mathcal{T} \; \text{exp}\left( -i \sum_{j=1}^n \int d^dxdt \lambda_j \varepsilon_j(t)\hat{Q}_j(t)f_j(\mbf{x} - \mbf{X}_i(t))\hat{\phi}(x) \right).
\label{eqn:prelimU}
\ee

Evolving the detector-field system using $\hat{U}$, the state of the detectors at time $\tau$ can be found by tracing out the field

\be
\rho_{D}(\tau) = \text{Tr}_{\phi}\left[ \ket{\Psi(\tau)}\bra{\Psi(\tau)} \right].
\ee

In \cite{Simidzija:2017jpo}, the authors compute $\hat{U}$ perturbatively up to second order, with arbitrary switching functions $\varepsilon_j(t)$, in their study of entanglement harvesting using two detectors. In \cite{Simidzija:2017kty}, however, a non-perturbative analysis of entanglement harvesting is performed by choosing Dirac-delta switching functions, so that the detectors couple with the scalar field at a discrete instant of time. Practically, this choice of switching function is a realistic assumption in the current imaging context given that the best modern photodetectors, such as Superconducting Nanowire Single-Photon Detectors \cite{0953-2048-25-6-063001}, have gating times on the order of picoseconds (so detectors separated by millimetres lie outside each others' lightcones). We will hence adopt a non-perturbative approach.

\section{Non-perturbative expression for $\hat{U}$}

Assume all detectors couple to the field at time $\tau$, so that $\varepsilon_j(t) = \delta(t - \tau)$. This allows us to factorise $\hat{U}$ into components acting on each detector individually using the BCH formula \cite{hall2003lie}

\be
\hat{U} = \text{exp}\left( \sum_{j=1}^n \hat{Q}_j(\tau) \otimes \hat{Y}_j(\tau) \right) = \prod_{j=1}^n \text{exp}\left( \hat{Q}_j(\tau) \otimes \hat{Y}_j(\tau) \right),
\label{eqn:Usimplifiedmore}
\ee

\noindent
where

\be
\hat{Y}_j(\tau) = -i \lambda_j \int d^dp \left[ F_j(\mbf{p},X_j(\tau))\hat{a}^{\dagger}_{\mbf{p}} + c.c. \right],
\label{eqn:Y}
\ee

\be
F_j(\mbf{p},X_j(\tau)) = \frac{\tilde{f}_j(\mbf{p})}{(2\pi)^d\sqrt{2 E_{\mbf{p}}}}e^{i(E_{\mbf{p}}\tau - \mbf{p} \cdot \mbf{X}_j(\tau))},
\ee

\noindent
with $\tilde{f}_j$ the Fourier transform of $f_j$. The operator $\hat{Q}_j(\tau)$ is an involution (i.e. $\hat{Q}^2_j(\tau) = \hat{\mbf{1}}_j$), and hence each exponential factor in Eq. (\ref{eqn:Usimplifiedmore}) can be expressed in terms of hyperbolic functions using their Taylor series expansions 

\bea
\text{exp}\left( \hat{Q}_j(\tau) \otimes \hat{Y}_j(\tau) \right) & = & \hat{\mbf{1}}_j \otimes \text{cosh}\left( \hat{Y}_j(\tau) \right) \nonumber \\
& & + \; \hat{Q}_j(\tau) \otimes \text{sinh}\left( \hat{Y}_j(\tau) \right).
\label{eqn:expQY}
\eea

Next, defining $\hat{X}^{i_s}_{j_s}(\tau) = \frac{1}{2}\left( e^{\hat{Y}_{j_s}(\tau)} + i_s e^{-\hat{Y}_{j_s}(\tau)} \right)$ and

\be
\hat{\delta}^{i_s}_{j_s}(\tau) = \begin{cases}
\hat{\mbf{1}}_{j_s} & \text{if } i_s = 1 \\
\hat{Q}_{j_s}(\tau) & \text{if } i_s = -1,
\end{cases}
\label{eqn:defdelta}
\ee

\noindent
Eq. (\ref{eqn:expQY}) can be recast as

\be
\text{exp}\left( \hat{Q}_j(\tau) \otimes \hat{Y}_j(\tau) \right) = \sum_{i_s \in\{ -1, 1 \}} \left( \hat{\delta}^{i_s}_j(\tau) \otimes \hat{X}^{i_s}_{j}(\tau) \right).
\label{eqn:esingle}
\ee

Suppressing the $\tau$ dependence from here onwards and using Eq. (\ref{eqn:esingle}) to rewrite Eq. (\ref{eqn:Usimplifiedmore}) gives an exact expression for the time-evolution operator

\be
\hat{U} = \sum_{\bar{j} \in \{ -1,1 \}^n} \hat{\delta}^{j_1}_1 \otimes \hat{\delta}^{j_2}_2 \cdots \hat{\delta}^{j_n}_n \otimes \hat{X}_{(\bar{j})},
\label{eqn:finalU}
\ee

\noindent
where $\hat{X}_{(\bar{j})} = \prod_{m=1}^n \hat{X}^{j_m}_{m} $, with $\bar{j} = (j_1,j_2,\cdots,j_n) \in \{ -1,1 \}^n$. Finally, since all the $\hat{Y}_j$ operators commute (they all act on the scalar field, at the same time), $\hat{X}_{(\bar{j})}$ can be expressed as

\be
\hat{X}_{(\bar{j})} = \frac{1}{2^n} \sum_{\bar{k} \in \{ -1,1 \}^n} \tilde{\delta}_{j_1,k_1}\tilde{\delta}_{j_2,k_2} \cdots \tilde{\delta}_{j_n,k_n} e^{\sum_{t=1}^n k_t \hat{Y}_t},
\label{eqn:finalX}
\ee

\noindent
where 

\be
\tilde{\delta}_{j_s,k_s} = \begin{cases}
1 & \text{if } k_s = 1 \\
j_s & \text{if } k_s = -1.
\end{cases}
\ee

\section{SPDC state model}
\label{sec:newspdcmodel}

To create an entangled photon state in typical quantum optics experiments, a coherent source of light such as a laser is directed onto a non-linear crystal, which creates an indeterminate number of pairs of entangled photons by way of Spontaneous Parametric Down Conversion (SPDC). To model such a situation, we begin with the following quantum field theory version of a coherent state categorized by a coherent amplitude $\alpha(\mbf{q})$ \cite{Simidzija:2017kty}, \cite{PhysRevA.88.032305}

\bea
\ket{\alpha(\mbf{q})} & := & \hat{D}_{\alpha(\mbf{q})} \ket{0} \nonumber \\
& := & \text{exp}\left( \int d^dq \left[ \alpha(\mbf{q})\hat{a}^{\dagger}_{\mbf{q}} - \alpha^*(\mbf{q})\hat{a}_{\mbf{q}} \right] \right) \ket{0}.
\label{eqn:coherentstate}
\eea

A pump photon with momentum $\mbf{q}$ from the coherent state has some probability of creating a pair of photons with momenta $\mbf{p}$ and $\mbf{k}$. This probability, along with the nature of the entanglement between the two output photons, is encapsulated in the quantity $\chi(\mbf{p},\mbf{k};\mbf{q})$, and this process is modeled by making the replacement

\be
\hat{a}^{\dagger}_{\mbf{q}} \to \int d^dp d^dk \chi(\mbf{p},\mbf{k};\mbf{q}) \hat{a}^{\dagger}_{\mbf{p}} \hat{a}^{\dagger}_{\mbf{k}},
\label{eqn:replacement}
\ee

\noindent
in Eq. \ref{eqn:coherentstate}. Note that this substitution necessarily involves post-selecting on the photons which undergo the down conversion process. We define $\hat{J}$ such that our SPDC state is

\bea
& & e^{\hat{J}} \ket{0} \nonumber \\
& & := \text{exp}\left( \int d^dq d^dp d^dk \left[ \alpha(\mbf{q})\chi(\mbf{p},\mbf{k};\mbf{q}) \hat{a}^{\dagger}_{\mbf{p}}\hat{a}^{\dagger}_{\mbf{k}} - c.c. \right] \right) \ket{0}. \nonumber \\
\label{eqn:almostfinalcoherentstate}
\eea

Some comments are in order. First, Eq. (\ref{eqn:almostfinalcoherentstate}) is invariant under exchange of the two photons, so $\chi(\mbf{p},\mbf{k};\mbf{q}) = \chi(\mbf{k},\mbf{p};\mbf{q})$. It also does not account for the annihilation of the pump photon: it merely creates two photons. Thus, the \textit{semiclassical approximation} is applicable given the extreme inefficiency of the SPDC process in practice. Finally, it is worth mentioning that since the SPDC process occurs in a non-linear crystal, Lorentz symmetry is consequently broken. Hence all our results need to be interpreted from the laboratory reference frame.

\section{Ghost imaging scheme}
\label{sec:ghostimagingscheme}

Let $\ket{\Psi(t_0)} = \ket{\mbf{E}} \otimes \ket{\mbf{K}}$ with $\ket{\mbf{K}} = e^{\hat{J}}\ket{0}$ be the initial state of the detector-field system. Evolving the system and tracing out the field at time $t > \tau$ gives the state of the detectors

\bea
\rho_D & = & \text{Tr}_{\phi} \left( \hat{U}\ket{\Psi(t_0)}\bra{\Psi(t_0)}\hat{U}^{\dagger} \right) \nonumber \\
& = & \sum_{\bar{j},\bar{l}} \hat{\delta}^{j_1}_1 \cdots \hat{\delta}^{j_n}_n \ket{\mbf{E}} \bra{\mbf{E}} \hat{\delta}^{l_1}_1 \cdots \hat{\delta}^{l_n}_n \mathscr{G}(\bar{l},\bar{j}),
\label{eqn:finalcoherentrho}
\eea

\noindent
with $\mathscr{G}(\bar{l},\bar{j}) = \bra{0} e^{-\hat{J}} \hat{X}^{\dagger}_{(\bar{l})} \hat{X}_{(\bar{j})} e^{\hat{J}}\ket{0}$. An expression for $\mathscr{G}$ can be found (see section \ref{subsec:calculateG}), and in what follows below, $F(\mbf{p}) = \sum_t (k_t - m_t) \lambda_t F_t(\mbf{p},X_t(\tau))$ and $\mathbb{X}(\mbf{p}_1,\mbf{p}_2) = 2(2\pi)^d \int d^dq \alpha(\mbf{q})\chi(\mbf{p}_1,\mbf{p}_2;\mbf{q})$

\bea
\mathscr{G}(\bar{l},\bar{j}) & = & \frac{1}{2^{2n}}\sum_{\bar{k},\bar{m} \in \{ -1,1 \}^n} \left( \prod_{s,v = 1}^n \tilde{\delta}_{l_{v},m_{v}}\tilde{\delta}_{j_{s},k_{s}} \right) \nonumber \\
& & \times \text{exp} \left( -\frac{1}{2} \int d^dp \left| \beta(\mbf{p}) \right|^2 \right),
\label{eqn:Gjkalmost}
\eea

\noindent
with

\begin{widetext}

\bea
\beta(\mbf{p}) = -i \int d^dp_0 [F(\mbf{p}_0)(\delta^{(d)}(\mbf{p}_0 - \mbf{p}) + K_{e}(\mbf{p}_0,\mbf{p})) + F^*(\mbf{p}_0)(\mathbb{X}(\mbf{p}_0,\mbf{p}) + K_{o}(\mbf{p}_0,\mbf{p}))],
\label{eqn:formulaforbeta}
\eea

\be
K_{e}(\mbf{p}_0,\mbf{p}) = \sum_{\substack{k = 2 \\ k \; \text{even}}} K_{e,k}(\mbf{p}_0,\mbf{p}) = \sum_{\substack{k = 2 \\ k \; \text{even}}} \frac{1}{k!} \int d^dp_1 \cdots d^dp_{k-1} \mathbb{X}^*(\mbf{p}_0,\mbf{p}_1)\mathbb{X}(\mbf{p}_1,\mbf{p}_2) \cdots \mathbb{X}(\mbf{p}_{k-1},\mbf{p}),
\label{eqn:formulaKe}
\ee

\be
K_{o}(\mbf{p}_0,\mbf{p}) = \sum_{\substack{k = 3 \\ k \; \text{odd}}}K_{o,k}(\mbf{p}_0,\mbf{p}) = \sum_{\substack{k = 3 \\ k \; \text{odd}}}\frac{1}{k!} \int d^dp_1 \cdots d^dp_{k-1} \mathbb{X}(\mbf{p}_0,\mbf{p}_1)\mathbb{X}^*(\mbf{p}_1,\mbf{p}_2) \cdots \mathbb{X}(\mbf{p}_{k-1},\mbf{p}).
\label{eqn:formulaKo}
\ee

\end{widetext}

As a sanity check, it can easily be seen that $\rho_D$ satisfies the properties of a density matrix. The quantities $\chi$ and $\alpha$ can freely be chosen to model specific dynamics of the SPDC process, and are given a detailed treatment in many works (such as \cite{PhysRevA.77.033808}). A perfect laser beam's angular spectrum, $\alpha(\mbf{q})$, is typically described by a Gaussian distribution. However, if the length scales in an experiment are much larger than the wavelength of the laser, the plane-wave approximation is apt, in which case $\alpha$ can be taken to be a Dirac-delta distribution, $\alpha(\mbf{q}) = \delta^{(d)}(\mbf{q} - \mbf{p}_p)$ for some known constant $\mbf{p}_p$, the momentum vector of the monochromatic photons emanating from the coherent pump source. We make this assumption here for ease of calculation. Furthermore, assuming a perfect phase-matching condition in the non-linear crystal in which perfect momentum conservation occurs, we can choose $\chi(\mbf{p}_1,\mbf{p}_2;\mbf{q}) = \delta^{(d)}(\mbf{p}_1 + \mbf{p}_2 - \mbf{q})$, in which case $\beta$ simplifies to

\bea
\beta(\mbf{p}) = -i\left( F(\mbf{p})\text{cosh}(2(2\pi)^d) + F^*(\mbf{p}_p - \mbf{p})\text{sinh}(2(2\pi)^d) \right). \nonumber \\
\label{eqn:simplifiedbeta}
\eea

Note that the perfect momentum conservation condition forces the total photon momentum in the plane transverse to $\mbf{p}_p$ to be zero: if a photon passes through point $x_T$ in the transverse plane, the other passes through point $-x_T$. Furthermore, in the simulations to follow, $\beta$ has been rescaled to emphasise the resultant correlations.

\subsection{2-detector inertial ghost imaging}
\label{sec:2coherentUDWexample}

Suppose that we have two parties, Alice and Bob, who are each given a detector, labelled 1 and 2 respectively. Alice prepares her detector in the excited state $\ket{e}_1$, which is interpreted as a binary image consisting a single white `on' pixel; Bob, without loss of generality, prepares his detector in the ground state $\ket{g}_2$. If the detectors are judiciously placed in space and if at time $\tau$ a photon from the SPDC state interacts with Alice's detector, given the perfect momentum conservation assumed earlier, there is a reasonable chance that the other photon interacted with Bob's detector (although this is definitely not certain, since the vacuum excitation probability of any detector alone is non-zero).

Next, we adopt a post-selection scheme in which Alice post-selects on the ground state (if we were to adopt a scheme entailing Alice post-selecting on `seeing a change' in the state of her detector, it can easily be shown that in the current single pair case, such a scheme prevents Bob from ascertaining whether Alice chose a white or black image). After post-selecting, Alice communicates the result to Bob via a classical channel: if she measures the ground state for her detector, she instructs Bob to measure his detector and keep the result; if she observes the excited state, Alice and Bob discard their detectors. Either way, they then re-prepare the initial state and run the experiment again. By repeating this experiment multiple times, Bob can infer the probability of Alice having chosen the ground or excited state.

So, with $\ket{\mbf{E}} = \ket{e}_1\ket{g}_2$, the state at some time $t > \tau$ is

\be
\rho_{D,AB} = \sum_{j_1,j_2,l_1,l_2} \hat{\delta}^{j_1}_1 \hat{\delta}^{j_2}_2 \ket{e}_1\ket{g}_2\bra{e}_1\bra{g}_2 \hat{\delta}^{l_1}_1 \hat{\delta}^{l_2}_2 \mathscr{G}(l_1,l_2;j_1,j_2).
\label{eqn:statebeforeAps}
\ee

Alice post-selects on $\ket{g}_1\bra{g}_1$ and communicates her result to Bob, whose state is obtained after tracing out Alice's detector after post-selection

\bea
\rho_{D,B} & = & \mathcal{K} \; \text{Tr}_{1}\left[ \rho_{D,AB} \ket{g}_1\bra{g}_1 \right] \nonumber \\
& = & \mathcal{K} \sum_{j_2,l_2} \hat{\delta}^{j_2}_2 \ket{g}_2\bra{g}_2 \hat{\delta}^{l_2}_2 \mathscr{G}(-1,l_2;-1,j_2),
\label{eqn:Bob'sstate2detector}
\eea

\noindent
with $\mathcal{K}$ such that $\text{Tr}(\rho_{D,B}) = 1$. Bob can thereafter use this state to reconstruct a ghost image by taking a convex sum of the two possible detector states, with the weightings being the probabilities themselves, i.e.

\be
\text{Ghost image} = P(g_2) \times \blacksquare + P(e_2) \times \square,
\label{eqn:2detectorimage}
\ee

\noindent
where $P(g_2)/P(e_2)$ is the probability of Bob observing his detector in the ground/excited state.

\begin{figure}[H]
\centering
\includegraphics[width=0.45\textwidth]{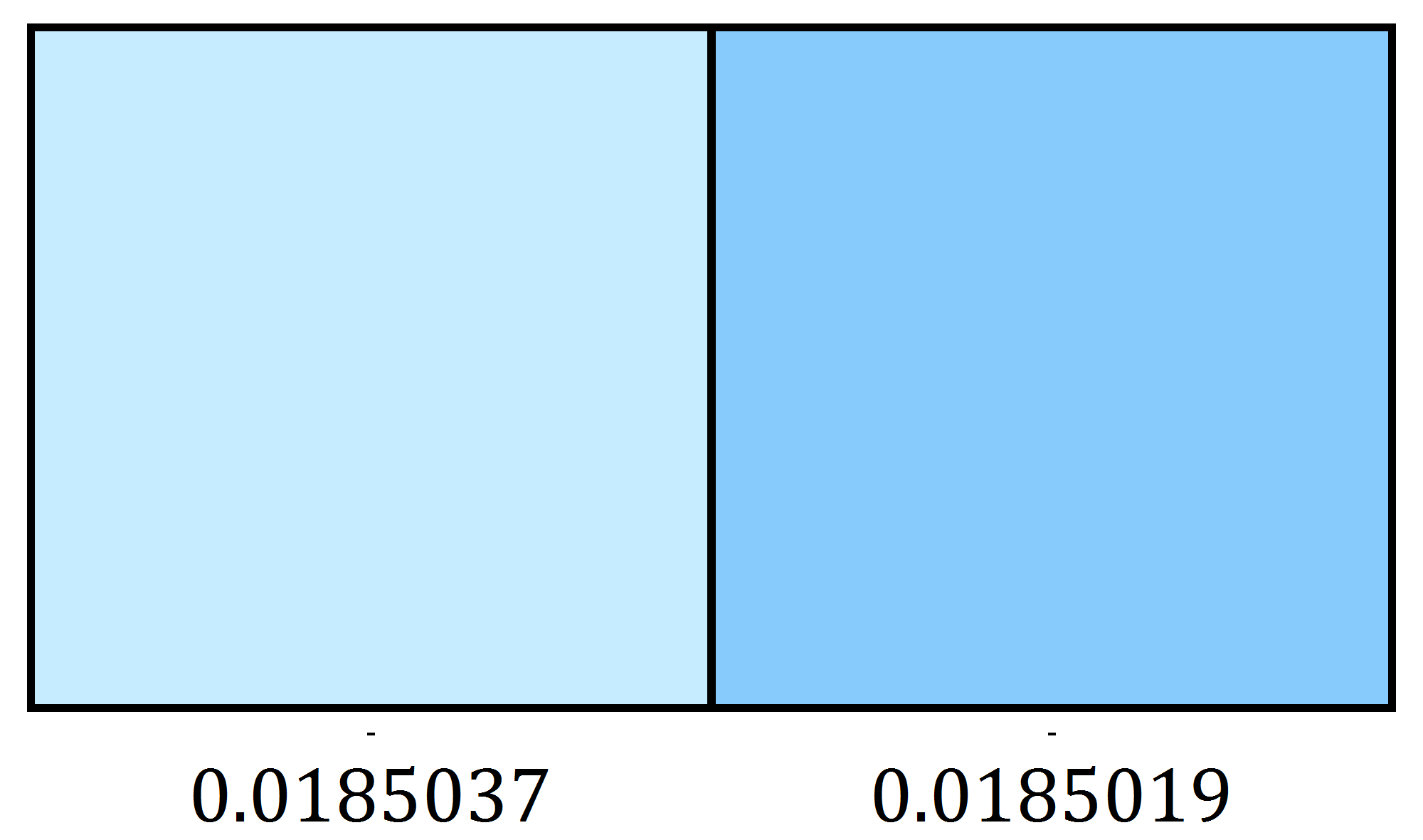}
\caption{Simulation of the two possible false-colour ghost images (re-scaled, for clarity) that Bob obtains if Alice initially prepares her detector in the ground (left) or excited (right) state. The pixel intensity, as per Eq. (\ref{eqn:2detectorimage}), is given below each image, with the parameters used in this simulation given in section \ref{subsec:parameters}, Table \ref{tab:parameter2detinertial}}
\label{fig:2detectorsimulation}
\end{figure}

In this scheme, assuming that since Alice prepares one of two initial detector states (and not a superposition of detector states), it is only crucial that the grey-level of the resultant ghost image in one case differs from the grey-level in the other case: a `brighter' ghost image will correspond with one possibility, and a `darker' result the other. Fig. \ref{fig:2detectorsimulation} gives a simulation of the reconstructed, re-scaled images Bob could obtain.

The pixel intensity indeed differs between the two possible outcomes: if Bob reconstructs an image with a pixel intensity corresponding to the right pixel in Fig. \ref{fig:2detectorsimulation}, he can conclude, better than a guess, that Alice chose the excited, `on' pixel as her ghost image, etc.

The contrast between the two possible images, although well within the numerical error, is low. This could, in part, be chalked up to the Dirac-delta coupling between the detectors and the initial field state. Studying the present scheme in the perturbative regime, with finite switching functions, would prove interesting since it is likely we would see an increase in the contrast for switching functions with larger supports.

It is, as yet, unclear how to satisfactorily define an entanglement measure for systems of higher dimension. However, in the current $2 \times 2$ dimensional case,  a measure such as the negativity $\mathcal{N}$ of a state \cite{PhysRevA.65.032314} can be used to investigate the strength of the correlations between Alice and Bob's detectors. $\mathcal{N}$ is defined by

\be
\mathcal{N}(\rho_{AB}) = - \sum_i{}^{\bf{'}} \text{E}^{T_B}_{AB,i},
\label{eqn:negativity}
\ee

\noindent
where $\text{E}^{T_B}_{AB,i}$ are the eigenvalues of $\rho_{AB}^{T_B}$, the partial transpose of $\rho_{AB}$ with respect to subsystem $B$, and the prime on the sum indicates summation over only \textit{negative} eigenvalues. $\mathcal{N}$ is zero iff the state $\rho_{AB}$ is separable. Interestingly, the negativity of the state in Eq. (\ref{eqn:statebeforeAps}), with the parameters of Table \ref{tab:parameter2detinertial}, is $0$ (at least within numerical error). Despite this, there is a clear correlation between Alice's initial state and the ghost image pixel in Fig. \ref{fig:2detectorsimulation}. This raises interesting questions as to its source. Firstly, it's certainly true that forgoing the perhaps unrealistic choices for $\alpha$, $\chi$, and the switching function (which can be interpreted as the gating profile of the detectors in the laboratory) and thereby choosing better pump beam profiles and/or phase matching conditions, could give rise to a non-zero negativity and stronger correlations between the detectors. However, it is also well known that separability is not sufficient to ensure the absence of quantum correlations in a state: in particular, quantum correlations can be present in mixed, separable states, a phenomenon described by the notion of \textit{quantum discord} \cite{articlediscord}. Eq. (\ref{eqn:statebeforeAps}) is both mixed and separable. Hence, this model may be accounting for quantum ghost imaging by way of stronger than classical correlations, without explicit entanglement existing between the detectors.

\subsection{4-detector inertial ghost imaging}
\label{sec:4coherentUDWexample}

In the 2 detector simulation above, the information content of the image Alice initially encodes in her single detector, namely 1 bit, equals the information content of the message sent over the classical channel to Bob. To demonstrate ghost imaging with an image containing more than 1 bit of information yet still employing the 1 bit channel, suppose that Alice and Bob are each given a pair of detectors: Alice's labelled 1 and 2, Bob's 3 and 4. Alice prepares her detectors in the state $\ket{e}_1\ket{g}_2$, a two-pixel binary image with pixel 1 white (`on') and pixel 2 black (`off'); Bob, without loss of generality, prepares both of his detectors in their ground states. Assuming the same scheme as above to find $\rho_{D,B}$ and having Bob measure this state to determine the statistics, his reconstructed image is given by

\bea
\text{Ghost image} & = & P(g_3,g_4) \cdot \blacksquare \blacksquare + P(g_3,e_4) \cdot \blacksquare \square \nonumber \\
& & + P(e_3,g_4) \cdot \square \blacksquare + P(e_3,e_4) \cdot \square \square.
\label{eqn:4detectorimage}
\eea

\begin{figure}[H]
\centering
\includegraphics[width=0.26\textwidth]{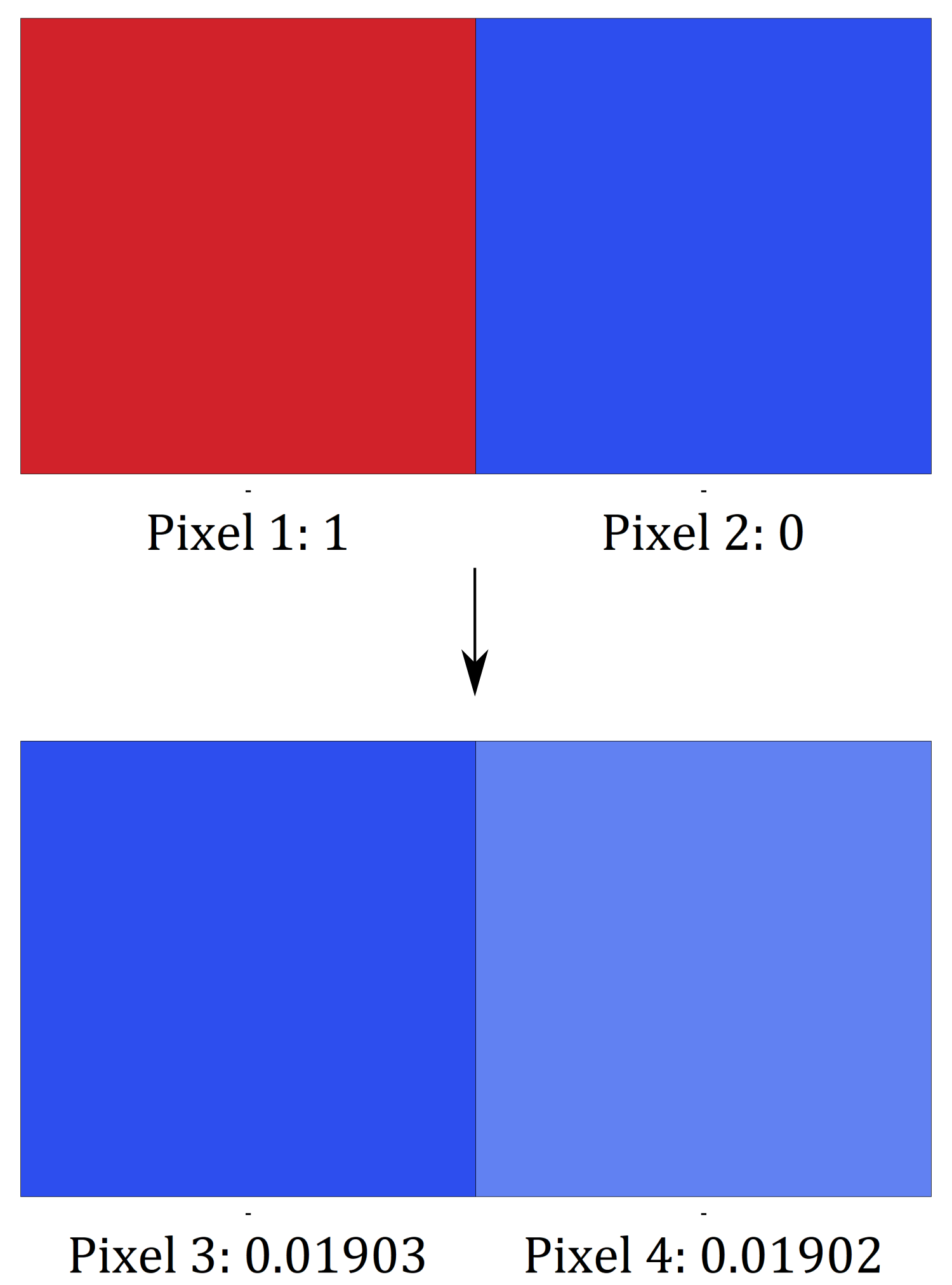}
\caption{False-colour, re-scaled simulation of the image Bob reconstructs (bottom), assuming that Alice initially prepares a `white-black' image (top). The parameters employed are given in section \ref{subsec:parameters}, Table \ref{tab:parameters4det}}
\label{fig:ddetectorsimulation}
\end{figure}

Fig. \ref{fig:ddetectorsimulation} gives a simulation of the two-pixel image Bob reconstructs given Alice's initial `excited-ground' image.

This simulation demonstrates true ghost imaging with a more complicated image. Sending images comprised of more pixels is possible, but the contrast between the possible ghost images Bob recreates seems to decrease markedly with the pixel number.

\section{Non-inertial ghost image}
\label{subsec:2detectornoninertial}

Non-inertial motion has a non-trivial effect on entanglement \cite{PhysRevD.85.061701}. Hence, if Alice's detector were to start uniformly accelerating, the correlations between the two detectors, and hence the image Bob reconstructs, would be affected. Furthermore, the ghost image will change depending on the time instant $\tau$ at which the detectors couple to the scalar field. The UDW formalism can account for such a non-inertial situation.

\begin{figure}[h]
\centering
\includegraphics[width=0.45\textwidth]{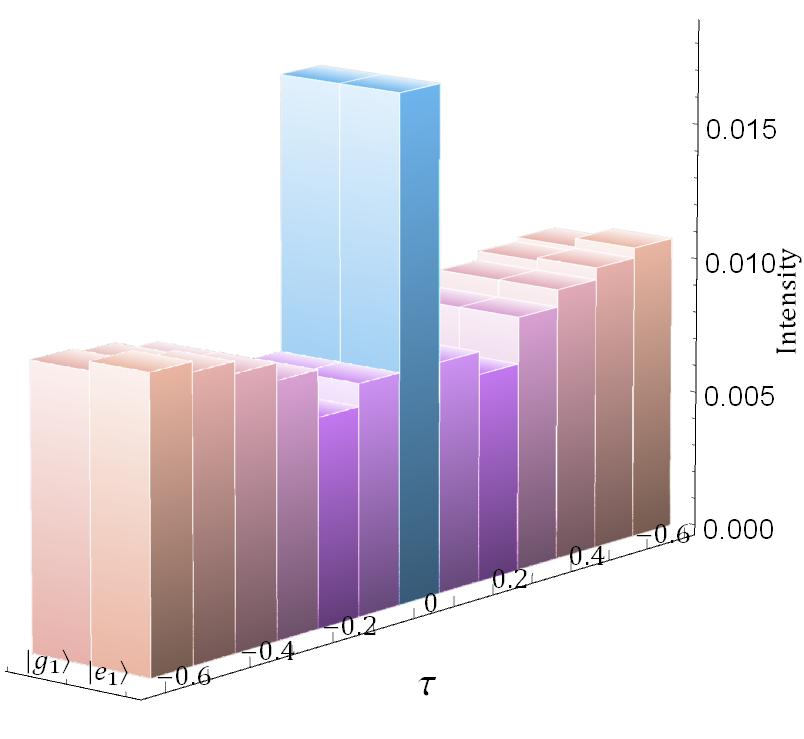}
\caption{Simulation of the ghost image pixel intensities that Bob obtains if the detectors couple to the field at time $\tau$, if Alice initially prepares her detector in either the ground or excited state. The parameters employed are given in Table \ref{tab:parameters2detnoninert}. Note the more prominent difference between the columns for larger values of $\tau$}
\label{fig:acceleratedetectors}
\end{figure}

Assume the same setup as before, with Bob's detector stationary with respect to the laboratory frame. Alice's detector, positioned such that it is temporarily at rest when $\tau = 0$ (corresponding to Fig. (\ref{fig:2detectorsimulation}), when Alice's detector is the same distance from the origin as Bob's), is uniformly accelerating in the $x$ direction with a proper acceleration $a$. The coordinates of Alice's detectors $(T,X,Y,Z)$, measured from the laboratory frame, are then given by the well-known Rindler coordinates

\bea
T = x \; \text{sinh}(a t), X = x \; \text{cosh}(a t), Y = y, Z = z,
\label{eqn:hyperbolicmotion}
\eea

\noindent
where $x = \frac{1}{a}$ is constant. Fig \ref{fig:acceleratedetectors} gives the reconstructed ghost image pixel intensities Bob would obtain for a small sample of various possible detector coupling instants $\tau$, if Alice initially prepares either a black (ground) or white (excited) pixel for her detector.

\begin{figure}[h]
\centering
\includegraphics[width=0.48\textwidth]{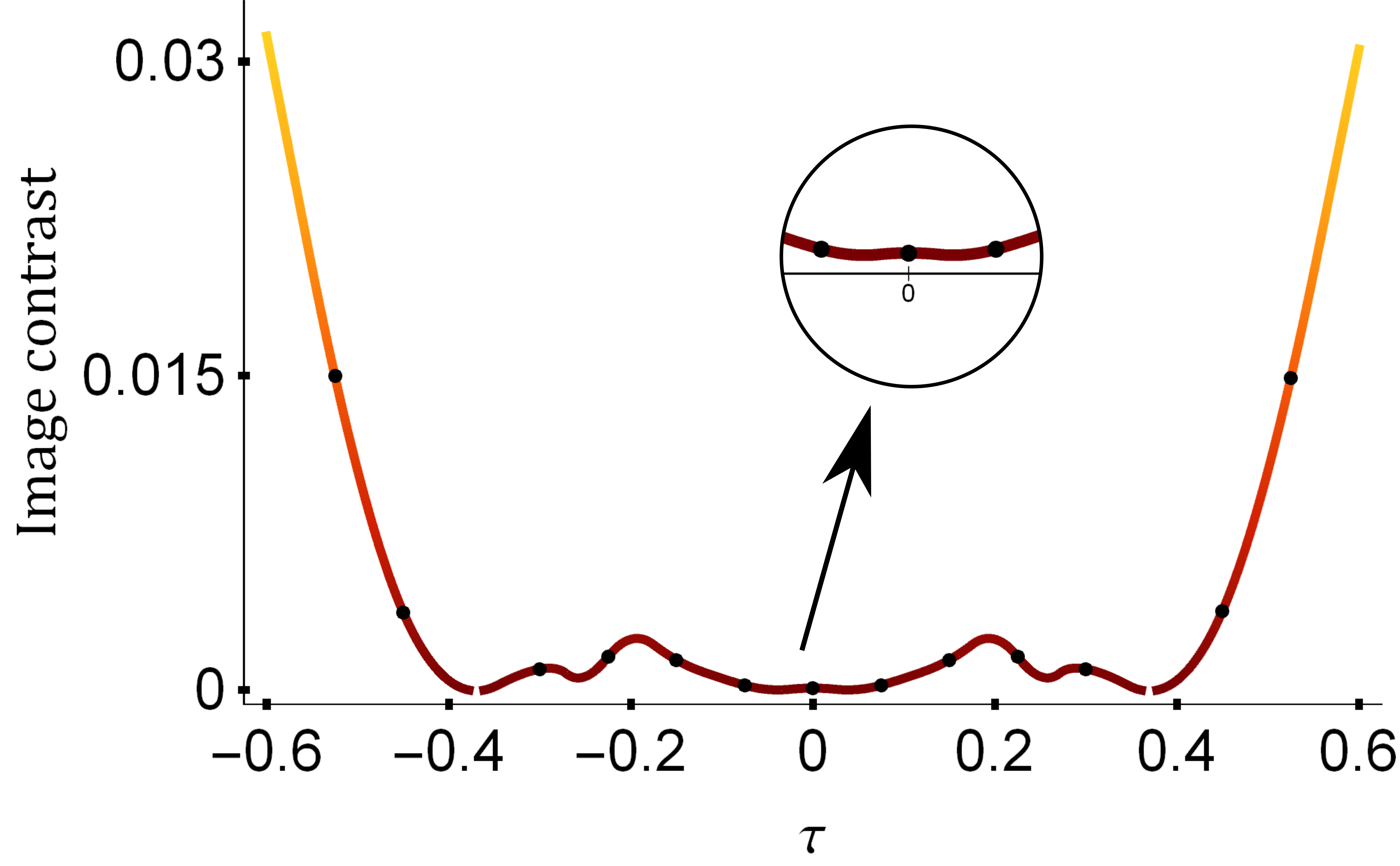}
\caption{Plot of the contrast between the two possible reconstructed image intensity values of Fig. \ref{fig:acceleratedetectors} as a function of $\tau$}
\label{fig:accelerateddetectorcontrast}
\end{figure}

The contrast between the two possible reconstructed ghost images (defined as the pixel intensity difference over the average pixel intensity) in Fig. \ref{fig:acceleratedetectors}, although quite small, does exist. Therefore, the ghost imaging scheme presented in this paper can be used, in theory, to communicate between non-inertial frames. However, the degree of contrast between the two possible outcomes (the `size' of the difference between the columns in Fig. \ref{fig:acceleratedetectors}) follows the unexpected, non-linear behaviour shown in Fig. \ref{fig:accelerateddetectorcontrast} for the $\tau$ values sampled in Fig. \ref{fig:acceleratedetectors}. At least intuitively, one would expect the contrast to decrease as $|\tau|$ decreases since a photon passing through the point $x_T$, in the plane transverse to the pump momentum vector $\mbf{p}_p$, occupied by Bob's detector at $\tau = 0$, is correlated with a photon passing through point $-x_T$, which is where Alice's detector lies when $\tau = 0$. However, the contrast is more evident at larger $|\tau|$ values. It would be interesting to investigate whether this counter-intuitive behaviour is merely a relic of the parameters chosen for the numerical simulation of the $\mathscr{G}$ values, or a more general phenomenon. For example, one could choose as the initial detector state a suitably `pre-entangled' state, which would eliminate any complications arising from the SPDC state; the situation may well change for smoother, more `well-behaved' switching functions (detector gating profiles).

\section{Discussion \& Conclusion}
\label{sec:conclusion}

We have here presented a model of the imaging process in quantum optics within a framework accommodating of relativity. The Unruh-DeWitt detectors, modelling the pixels, couple to the field to be probed by means of a Dirac-delta switching function, with the formulae leaving the detector spatial profiles general.

We presented a state modelling the SPDC process in quantum optics, as well as a scheme to perform quantum ghost imaging using both the UDW detectors and the SPDC state. The SPDC field state is quite general (only the semi-classical approximation was made, which is widely used in practice) and is clearly normalizable. Previous studies have usually concentrated on harvesting entanglement from the vacuum, and it could be argued that the vacuum is more often the exception rather than the norm in the laboratory. Developing a model for an entangled, non-empty state such as that in Eq. (\ref{eqn:almostfinalcoherentstate}), may prove fruitful. However, this SPDC state results in a relatively unwieldy solution for the density matrix of the coupled UDW detectors once it is traced out. Indeed, Eqs. (\ref{eqn:formulaKe}) and (\ref{eqn:formulaKo}) satisfy a recursion relation which could be used to find a closed-form expression for $\beta$ (one can use it to define a multi-dimensional integro-differential equation which a solution must satisfy), although it is not clear whether such an expression exists. Furthermore, although convenient, the assumption of a Dirac-delta switching function in the interaction Hamiltonian is somewhat idealistic: most photon counting modules employed in the laboratory have gating times resembling square wave functions. A finite switching function, coupling the detectors to the field for longer periods of time, may increase the correlations between the detectors, and hence the image contrast (particularly with reference to Fig. \ref{fig:acceleratedetectors}). Such an analysis would need to be perturbative in nature, but the effort would be well worth it.

Next, we presented numerical simulations of the proposed scheme within an inertial setting, given in Figs. \ref{fig:2detectorsimulation} and \ref{fig:ddetectorsimulation}, which confirms the potential of using this scheme to describe the imaging process. In the single-pixel case, we found that, interestingly, the negativity of the post-selected state between Alice and Bob in this simulation was 0, implying no entanglement between the parties. However, Bob could still infer, better than a simple guess, the state Alice initially prepared. It will hence be interesting to investigate the exact nature of the weak correlation between the detectors. Later work may also well look into non-binary ghost images, which would require UDW detectors with more than 2 possible states.

Finally, the model was employed to a uniformly accelerating detector. Although indeed present, the contrast between the different potential reconstructed images is extremely small. This is perhaps unsurprising, given the relatively small amount of entanglement harvested from the quantum vacuum by detectors coupled to it \cite{PhysRevD.92.064042}. Understanding the degree of contrast between potential outcomes is important to consider and improve upon if such studies are ever to be observed. Indeed, quantum ghost imaging could be seen as a form of secret sharing (the secret being the image itself), and the relaying of entangled qubits via satellites is becoming increasingly important \cite{1263786}. Furthermore, it would be interesting to consider imaging in non-trivial backgrounds, since, for example, rotating black holes alter the orbital angular momentum (OAM) of nearby photons \cite{tamburini2011twisting}, and OAM is increasingly showing use as an information-carrying degree of freedom in optical communication \cite{erhard2018twisted}.

\section{Acknowledgments}

N.B. acknowledges support from the South African CSIR IBS programme, and A.K. acknowledges support from the National Science and Engineering Research Council of Canada (NSERC).

\section{Supplementary Information}
\label{sec:SI}

\subsection{Calculation of $\mathscr{G}(\bar{l},\bar{j})$}
\label{subsec:calculateG}

Here we outline the calculation of $\mathscr{G}(\bar{l},\bar{j})$. Firstly, $\mathscr{G}$ can be recast in the following form

\begin{align}
\mathscr{G}(\bar{l},\bar{j}) = \frac{1}{2^{2n}}\sum_{\bar{k},\bar{m} \in \{ -1,1 \}^n} & \left( \prod_{s,v = 1}^n \tilde{\delta}_{l_{v},m_{v}} \tilde{\delta}_{j_{s},k_{s}} \right. \nonumber \\
& \left. \times \bra{0} e^{-\hat{J}} e^{\sum_t (k_t - m_t)\hat{Y}_t} e^{\hat{J}} \ket{0} \right).
\label{eqn:Gjkalmostalmost}
\end{align}

A braided corollary of the Baker-Campbell-Hausdorff formula \cite{hall2003lie},

\be
e^{\hat{A}}e^{\hat{B}}e^{-\hat{A}} = \text{exp} \left( \sum_{k=0}^{\infty} \frac{1}{k!} \underbrace{[ \hat{A}, [ \cdots , [ \hat{A} }_{k}, \hat{B} ] ] \cdots ] \right),
\label{eqn:BCHv2}
\ee

\noindent
allows $e^{-\hat{J}} e^{\sum_t (k_t - m_t)\hat{Y}_t} e^{\hat{J}}$ to be expressed in terms of nested commutators of $\hat{J}$ and $\sum_t (k_t - m_t)\hat{Y}_t$. Although the exponent in Eq. (\ref{eqn:BCHv2}) does not terminate in this case, a curious pattern for the even $k$ and odd $k$ terms does emerge

\begin{align}
& e^{-\hat{J}} e^{\sum_t (k_t - m_t)\hat{Y}_t} e^{\hat{J}} \nonumber \\
& = \text{exp}\left[ -i \int d^dp_0 d^dp ( [F(\mbf{p}_0)(\delta^{(d)}(\mbf{p}_0 - \mbf{p}) + K_{e}(\mbf{p}_0,\mbf{p})) \right. \nonumber \\
& \left. + F^*(\mbf{p}_0)(\mathbb{X}(\mbf{p}_0,\mbf{p}) + K_{o}(\mbf{p}_0,\mbf{p}))] \hat{a}^{\dagger}_{\mbf{p}} + c.c. ) \right],
\label{eqn:finalexpnewmodel}
\end{align}

\noindent
with the other quantities as per section \ref{sec:newspdcmodel}. Eq. (\ref{eqn:finalexpnewmodel}) clearly has the form of a coherent state displacement operator $\hat{D}_{\beta(\mbf{p})} = \text{exp}\left( \int d^dp \left[ \beta(\mbf{p}) \hat{a}^{\dagger}_{\mbf{p}} - \text{c.c.} \right] \right)$. So, the argument in Appendix A of \cite{Simidzija:2017kty} follows verbatim with $\beta$ as per Eq. (\ref{eqn:formulaforbeta})

\be
\bra{0} e^{-\hat{J}} e^{\sum_t (k_t - m_t)\hat{Y}_t} e^{\hat{J}} \ket{0} = \text{exp} \left( -\frac{1}{2} \int d^dp \left| \beta(\mbf{p}) \right|^2 \right).
\ee

\subsection{Parameters of simulations}
\label{subsec:parameters}

The following tables give the parameters and functions used in the simulations. We chose $d = 3, m = 0, \sigma = 1/10, \mbf{p}_p = (0,0,-2\pi)$, and $f(x,y,z) = \mathscr{N}\text{exp}(-(x^2+y^2 + z^2)/2\sigma^2)$ throughout.

\begin{table}[h!]
  \begin{center}
    \caption{Table of parameters for 2 detector inertial example}
    \label{tab:parameter2detinertial}
    \begin{tabular}{l|c}
      \textbf{Parameter} & \textbf{Value} \\
      \hline
      $\lambda, \tau$ & $1, 0$ \\
      $X_1^{\mu}(\tau)$ & $(\tau, 1, 0, 0)$ \\
      $X_2^{\mu}(\tau)$ & $(\tau, -1, 0, 0)$ \\
    \end{tabular}
  \end{center}
\end{table}

\begin{table}[h!]
  \begin{center}
    \caption{Table of parameters for 4 detector inertial example}
    \label{tab:parameters4det}
    \begin{tabular}{l|c}
      \textbf{Parameter} & \textbf{Value} \\
      \hline
      $\lambda, \tau$ & $1, 0$ \\
      $r, \theta$ & $1, \pi/6$ \\
      $X_1^{\mu}(\tau)$ & $(\tau, r \cos(\theta), r \sin(\theta),0)$ \\
      $X_2^{\mu}(\tau)$ & $(\tau, r \cos(\theta), -r \sin(\theta),0)$ \\
      $X_3^{\mu}(\tau)$ & $(\tau, -r \cos(\theta), -r \sin(\theta),0)$ \\
      $X_4^{\mu}(\tau)$ & $(\tau, -r \cos(\theta), r \sin(\theta),0)$ \\
    \end{tabular}
  \end{center}
\end{table}

\begin{table}[h!]
  \begin{center}
    \caption{Table of parameters for 2 detector non-inertial example}
    \label{tab:parameters2detnoninert}
    \begin{tabular}{l|c}
      \textbf{Parameter} & \textbf{Value} \\
      \hline
      $\lambda$ & $1$ \\
      $r, \theta$ & $1$, $0$ \\
      $a$ & $\frac{1}{r \text{cos}(\theta)}$ \\
      $X_1^{\mu}(\tau)$ & $(\frac{1}{a}\text{sinh}(a \tau), \frac{1}{a}\text{cosh}(a \tau), r \sin(\theta),0)$ \\
      $X_2^{\mu}(\tau)$ & $(\tau, -\frac{1}{a}, r \sin(\theta),0)$ \\
    \end{tabular}
  \end{center}
\end{table}

\bibliographystyle{apsrev4-1}
\bibliography{References}

\end{document}